\documentclass[preprint,showpacs,preprintnumbers,amsmath,amssymb]{revtex4}
\usepackage{graphicx}
\usepackage{epsfig}
\usepackage{bm}
\usepackage{amsfonts}
\usepackage{subfigure}
\usepackage{multirow}
\usepackage{float}
\usepackage{color}
\usepackage{xcolor}
\usepackage{ulem}
\usepackage{todonotes}
\usepackage{makecell}
\usepackage{enumitem}
\usepackage{float}
\usepackage{amsmath}
\usepackage{hyperref}
\hypersetup{colorlinks=True,citecolor=blue}
\usepackage{amsfonts}
\usepackage{amssymb}
\usepackage{caption}
\usepackage{float}
\usepackage{bm}
\usepackage{graphicx}
\usepackage{color}
\usepackage{verbatim}
\usepackage{subfigure}
\usepackage{multirow}
\usepackage{array}
\def\thickhline{\noalign{\hrule height1.5pt}}

\usepackage{enumitem}
\usepackage{float}
\usepackage{amsmath}
\usepackage{hyperref}
\hypersetup{colorlinks=True,citecolor=blue}
\usepackage{amsfonts}
\usepackage{amssymb}
\usepackage{caption}
\usepackage{float}
\usepackage{bm}
\usepackage{graphicx}
\usepackage{color}
\usepackage{verbatim}
\usepackage{subfigure}
\usepackage{multirow}

\def\beq{\begin{equation}}
\def\eeq{\end{equation}}
\def\ber{\begin{eqnarray}}
\def\eer{\end{eqnarray}}
\def\benu{\begin{enumerate}}
\def\eenu{\end{enumerate}}

\def\sq{\lower.25ex\hbox{\large$\Box$}}
\def \lleq {\lower0.9ex\hbox{ $\buildrel < \over \sim$} ~}
\def \ggeq {\lower0.9ex\hbox{ $\buildrel > \over \sim$} ~}

\def\prd{{Phys.\@ Rev.\@ D\ }}

\def\plb {{Phys.\@ Lett.\@ B\ }}

\begin{document}
\title{\textbf
{Reheating and relic gravitational waves as remedies for degeneracies of non-canonical natural inflation }}

\author{Karam Bahari$^{1}$, Soma Heydari$^{2,}$\footnote{Author to whom any correspondence should be addressed.\newline E-mail: s.heydari@uok.ac.ir} and Kayoomars Karami$^{2}$}

\affiliation{\small{
$^{1}$Physics Department, Faculty of Science, Razi University, Kermanshah, Iran\\
$^{2}$Department of Physics, University of Kurdistan, Pasdaran Street, P.O. Box 66177-15175, Sanandaj, Iran
}}
\begin{abstract}
Here, a natural non-canonical inflationary model based on a power-law Lagrangian is investigated. We analyze the scalar spectral index $n_{\rm s}$ and the tensor-to-scalar ratio $r$ of the model and identify their degeneracies with respect to the free parameters. Notably, $n_{\rm s}$ and $r$ show effective independence from the model parameters due to degeneracies in the slow-roll parameters that leads to unresolved parameter degeneracies. Employing the constraints on reheating parameters such as the reheating duration $N_{\rm{reh}}$, the reheating temperature $T_{\rm{reh}}$, and the equation of state parameter $\omega_{\rm{reh}}$, is found to be insufficient to fully break these degeneracies. However, the relic gravitational wave spectrum provides a way to break degeneracy with respect to the non-canonical parameter $\alpha$, degeneracy with respect to the potential parameter $f$ persist. Finally, we specify the allowed ranges for the inflationary duration $N$ and the parameter $\alpha$, in light of the latest observational data. These results highlight the role of relic gravitational waves in refining inflationary models and illustrate the challenges in fully resolving parameter degeneracies.

\end{abstract}
\maketitle
\section{Introduction}
The inflationary paradigm as a pivotal supplement to the Hot Big Bang (HBB) theory was established to expound foundational issues in early Universe physics. It is supported by observational evidence such as the Cosmic Microwave Background (CMB) and renders reliable framework to explain the horizon problem, the flatness of the Universe \cite{Guth:1981}, the lack of magnetic monopoles, and the source of large-scale structure \cite{Linde:1982}. During inflation, the Universe experiences accelerated expansion driven by a scalar field known as  inflaton which evolves on a potential \cite{Baumann:2009}. In this era, quantum fluctuations in the inflaton field generate  scalar and tensor perturbations. The scalar perturbations seed density fluctuations and the tensor perturbations produce relic gravitational waves (GWs) \cite{Starobinsky:1979}.

Commonly, a plateau-like potential in an inflationary model is responsible to set up a slow-roll phase. This phase is characterized by slow-roll parameters that remain much smaller than unity to confirm acceleration expansion during inflation. The slow-roll approximation fails when one of the slow-roll parameters approaches unity to denote the end of inflation. After inflation, the Universe transits into the radiation-dominated (RD) era through a reheating phase. During this phase, the inflaton oscillates around the potential minimum and releases its energy to reheat the cosmos and populate it with Standard Model particles \cite{Cook:2015}.

The reheating phase bridges inflation and the RD era. It is characterized by three parameters including the reheating duration $N_{\rm{re}}$, the reheating temperature $T_{\rm{re}}$, and the effective equation of state parameter $\omega_{\rm{re}}$.
These  parameters not only govern the thermal history of the Universe but also provide additional constraints on inflationary models. If theoretical limitations on $N_{\rm{re}}$, $T_{\rm{re}}$, and $\omega_{\rm{re}}$ merge with observational constraints from the CMB on the scalar spectral index $n_{\rm s}$ and the tensor-to-scalar ratio $r$, it is possible to tighten the parameter space of inflationary scenarios \cite{German:2023}. The  reheating constraints are investigated on mutated hilltop inflation in Ref. \cite{Yadav:2024} and quintessential inflation in Ref. \cite{de Haro:2024}. Reheating constraints also play an essential role for addressing degeneracies in inflationary models, which arise when different parameter values yield identical predictions for $n_{\rm s}$ and $r$ \cite{Mishra:2021}.  Thus, analyzing reheating parameters provides an additional way to break such degeneracies and refine the predictions of inflationary scenarios \cite{safaei:2024}.

In addition to the reheating,  relic GWs offer a powerful tool for breaking degeneracies \cite{safaei:2024-2} and further constraining inflationary model \cite{Haque:2025}. The density spectrum of relic GWs is proportional to $\omega_{\rm{re}}$ and subsequently is in proportion to model parameters \cite{sahni:1990}. The relic GWs propagate in the Universe with no interactions with radiation or matter. Hence, they convey pure information about the physics of the Universe in the various epochs. Moreover, the density spectrum of GWs could be traceable through the forthcoming GW detectors, if their frequencies lie inside the sensitivity zone of that detectors.

It is well-recognized that the predictions of the natural inflationary potential, in the standard model of inflation, cannot be consistent with the latest observational data of Planck 2018 \cite{akrami:2020}. To revive this potential, one can employ the modified gravity \cite{Teimoori:2021}, non-canonical models with power-law \cite{Heydari:2024} or field-dependent \cite{Solbi-a:2021} kinetic term. The non-canonical model of inflation with power-low Lagrangian, ${\cal L}(X,\phi) = X^{\alpha} - V(\phi)$,  is a well-known generalization of the standard model ($X$ denote the kinetic energy of the inflaton) \cite{Unnikrishnan:2013}. In this model, deviation from the canonical Lagrangian of the standard model of inflation is signified by parameter $\alpha$ \cite{Rezazadeh:2015}. Tuning of this parameter in the inflationary model enable us to revive the predictions of the steep potentials such as  \cite{Heydari:2024b}.

In this paper, we analyze the natural potential within the framework of non-canonical inflationary model containing the power-law Lagrangian. We examine a set of observational and theoretical constraints on this model including the Plank$+$BICEP/Keck 2018 data, reheating parameters bounds, and relic GWs limitations, focusing on constraining the model parameters, inflationary duration, and resolving potential degeneracies. To do this, in Section \ref{sec2}, we review the fundamental concepts of non-canonical inflationary model to constrain the observable parameters, $n_{\rm s}$ and $r$. In Section \ref{sec3}, we apply the reheating constraints to the model. In Section \ref{sec4}, we study the primordial GWs predicted by the model. In the end, the main conclusions are epitomized in Section \ref{sec5}.

\section{Non-canonical inflation}\label{sec2}
In this section, we focus on the following generic action
\begin{equation}\label{eq:action}
S=\int{\rm d}^{4}x \sqrt{-g} \left[\frac{R}{16\pi G}+{\cal L}(X,\phi)\right].
\end{equation}
Here $g$ and $R$ represent the determinant of the metric tensor $g_{\mu\nu}$ and the Ricci scalar, respectively. The function $\mathcal{L}(X,\phi)$ is the Lagrangian density which depends on the scalar field $\phi$ and the kinetic energy $X\equiv\frac{1}{2}g_{\mu\nu}\partial^\mu \phi \partial^\nu \phi$.
In the current work, a power-law function of the kinetic energy is used for the Lagrangian density as follows \cite{Heydari:2024a}
\begin{equation}\label{Lagrangian}
{\cal L}(X,\phi) = X\left(\frac{X}{M^{4}}\right)^{\alpha-1} - V(\phi),
\end{equation}
where $\alpha$ is the dimensionless non-canonical parameter, $V\left(\phi\right)$ is the inflationary potential and $M$ represents a mass scale parameter with the mass dimension. For the Lagrangian density (\ref{Lagrangian}), the energy density $\rho_{\phi}$ and pressure $p_{\phi}$ of the scalar field can be obtained as
\begin{equation}
\rho _\phi =\left ( \frac{\partial{\cal L} }{\partial X} \right )\left ( 2X \right )-{\cal L}=\left( {2\alpha - 1}
\right)X{\left( {\frac{X}{{{M^4}}}} \right)^{\alpha  - 1}} +
V(\phi) \label{eq:rho},
\end{equation}
\begin{equation}
{p_\phi } ={\cal L}=X{\left( {\frac{X}{{{M^4}}}} \right)^{\alpha  - 1}} - V(\phi ) .\label{eq:Lp}
\end{equation}
Here, we consider a spatially flat Friedmann-Robertson-Walker (FRW) universe given by the following line element
\begin{equation}
\label{eq:FRW}
{\rm d} {s^2} = {\rm d} {t^2} - {a^2}(t)\left( {{\rm d} {x^2} +
{\rm d} {y^2} + {\rm d} {z^2}} \right) ,
\end{equation}
where, $a(t)$ represents the scale factor as a function of cosmic time $t$.
Using this metric, the kinetic energy simplifies to  $X=\dot{\phi}^2/2$, where the dot denotes a derivative with respect to time.
By varying the  action (\ref{eq:action}) with respect to the metric (\ref{eq:FRW}), and using Eqs. (\ref{eq:rho}) and (\ref{eq:Lp}), the first and second Friedmann equations can be obtained as
\begin{equation}
H^{2} =\dfrac{1}{3M_p^2} \rho_{\phi} = \frac{1}{3M_p^2}\left[\left(2\alpha-1\right)X\left(\frac{X}{M^{4}}\right)^{\alpha-1} + V(\phi)\right] , \label{eq: FR-eqn1}
\end{equation}
\begin{equation}
\dot{H} = -\frac{1}{2M_p^2} (\rho _{\phi} +p_{\phi} ) = -\frac{1}{M_p^2}\alpha X\left(\frac{X}{M^4}\right)^{\alpha -1}.
\label{eq:FR-eqn2}
\end{equation}
Here, $M_p \equiv 1 / {\sqrt{8\pi G}}$ is the reduced Planck mass and $H \equiv \dot{a}/a$ is the Hubble parameter.
By varying the action (\ref{eq:action}) with respect to $\phi$, the equation governing the motion of scalar field $\phi$ takes the following form
\begin{equation}
\label{eq:KG}
\ddot \phi  + \frac{{3H\dot \phi }}{{2\alpha  - 1}}+ \left( {\frac{V'(\phi)}{{\alpha (2\alpha  - 1)}}}
\right){\left( {\frac{{2{M^4}}}{{{{\dot \phi }^2}}}} \right)^{\alpha- 1}} = 0,
\end{equation}
in which the prime represents the derivative with respect to the scalar field.
Note that in the case of $\alpha=1$, all of the equations discussed above reduce to their canonical form.
The slow-roll parameters are defined in terms of Hubble parameter as
\begin{equation}
\label{eq:H slow roll parameters}
\epsilon\equiv -\frac{\dot{H}}{H^2} ~ , ~ \eta \equiv \epsilon -\frac{\dot{\epsilon }}{2H\epsilon }.
\end{equation}
Using the slow-roll approximation, i.e. $ (\epsilon, \eta) \ll 1 $, the first Friedmann equation  (\ref{eq: FR-eqn1}) can be simplified  as
\begin{equation}
\label{eq:FR1-SR}
3 M_p^{2}H^2\simeq V(\phi).
\end{equation}
Furthermore, under the slow-roll approximation, the term $\ddot{\phi}$ in Eq. (\ref{eq:KG}) can be neglected and consequently using Eq. (\ref{eq:FR1-SR}), the equation of motion reduces to
\begin{equation}\label{eq:KG-SR}
\dot \phi  =  -
\theta \left[ {\left( {\frac{M_p}{{\sqrt 3 \alpha }}}
\right)\left( {\frac{{\theta {V}'(\phi )}}{{\sqrt {V(\phi )} }}}
\right){{\left( {2{M^4}} \right)}^{\alpha  - 1}}}
\right]^{\frac{1}{{2\alpha  - 1}}},
\end{equation}
in which $\theta=+1$, for $V'(\phi)>0 $ and $\theta=-1$, for $V'(\phi)<0$.
In the case of slow-roll approximation $ (\epsilon, \eta) \ll 1 $, and using Eqs. (\ref{eq:FR1-SR}) and (\ref{eq:KG-SR})  it can be shown that the slow-roll parameters (\ref{eq:H slow roll parameters}) can be written as functions of the inflationary potential and its derivative as follows \cite{Unnikrishnan:2012}
\begin{align}
&\epsilon \simeq \left [ \frac{1}{\alpha } \left ( \frac{3M^{4}}{V\left( \phi\right)} \right )^{\alpha -1}\left ( \frac{M_{p}{V}'\left( \phi\right) }{\sqrt{2}V\left(\phi\right)} \right )^{2\alpha }\right ]^{\frac{1}{2\alpha -1}} , \label{eq:epsilon V}\\
&\eta \simeq \left ( \frac{\alpha \epsilon }{2\alpha -1} \right )\left ( 2\frac{V\left ( \phi  \right ){V}''\left ( \phi  \right )}{{V}'\left ( \phi  \right )^2} -1 \right ). \label{eq:eta V}
\end{align}
Note that, for $\alpha=1$, Eqs. (\ref{eq:epsilon V}) and (\ref{eq:eta V}) reduce to $\epsilon=\epsilon_{V}$ and $\eta=\eta _{V}-\epsilon_{V}$, where $\epsilon_{V}\equiv\frac{M_p^2}{2}\left(\frac{V'}{V}\right)^2$ and $\eta _{V}\equiv M_p^2\left( \frac{V''}{V}\right)$ describe the potential slow-roll parameters obtained in the standard inflationary model.

According to \cite{Garriga:1999}, the power spectrum of scalar perturbation is given by
\begin{equation}\label{eq:Ps-SR}
{\cal P}_{s}=\frac{H^2}{8 \pi ^{2}M_p^{2}c_{s} \epsilon}\Big|_{c_{s}k=aH}.
\end{equation}
In this equation  $c_{_s}$ denotes the sound speed of the scalar perturbation defined as
\begin{equation}
\label{eq:cs-NC} c_s^2 \equiv \frac{{\partial {p_\phi }/\partial X}}{{\partial {\rho _\phi }/\partial X}} = \frac{{\partial {\cal L}(X,\phi )/\partial X}}{{\left( {2X} \right){\partial ^2}{\cal L}(X,\phi )/\partial {X^2} + \partial {\cal L}(X,\phi )/\partial X}}.
\end{equation}
From the Planck measurements, the amplitude of the scalar power spectrum is confined to ${\cal P}_{s} (k_{\ast })=2.1\times 10^{-9}$ at the pivot scale ($k_{*}=0.05~\text{Mpc}^{-1}$)  \cite{akrami:2020}. Using the Lagrangian (\ref{Lagrangian}), the square of the  sound speed (\ref{eq:cs-NC}) can be determined as
\begin{equation}\label{eq:cs2}
c_{s}^{2}=\frac{1}{2\alpha -1}.
\end{equation}
In order to avoid phantom and classical instabilities, the inequality $0\le c_{s}^{2}\le 1$ must be satisfied, consequently, from Eq. (\ref{eq:cs2}), we obtain  the condition $\alpha\ge 1$.
In both the canonical and non-canonical frameworks, the scalar spectral index $n_{s}$ is determined from the scalar power spectrum. It can also be expressed in terms of the slow-roll parameters as \cite{Garriga:1999}
\begin{equation}
n_{\rm s} - 1\equiv \frac{\mathrm{d} \ln {\cal P}_ {s}}{\mathrm{d} \ln k}=-4\epsilon+2\eta. \label{eq:inf_ns_SR}
\end{equation}
The observational data of Planck 2018 TT, TE, EE + LowE + Lensing + BK18 + BAO at 68\% and 95\% CL impose the following restrictions on the scalar spectral index $n_{\rm s}$ \cite{bk18,Paoleti:2022}
\begin{equation}\label{eq:ns95}
n_{\rm s} = 0.9653_{-\,0.0041\,-\,0.0083}^{+\,0.0041\,+\,0.0107} .
\end{equation}
As for the tensor perturbations, the tensor power spectrum is obtained as follows
\cite{Garriga:1999}
\begin{equation}\label{eq:Pt-SR}
{\cal P}_{t}=\frac{2H^2}{\pi ^{2}M_p^2}\Big|_{k=aH}.
\end{equation}
The tensor spectral index is defined as the logarithmic derivative of tensor power spectrum (\ref{eq:Pt-SR}) and it can be written in terms of first slow-roll parameter as
\begin{equation}
\label{eq:ntt}
n_{t}\equiv \frac{\mathrm{d} \ln {\cal P}_{t}}{\mathrm{d} \ln k}= -2 \epsilon.
\end{equation}
Using Eqs. (\ref{eq:Ps-SR}) and (\ref{eq:Pt-SR}), one can obtain the tensor-to-scalar ratio as
\begin{equation}
\label{eq:r}
r\equiv\frac{{\cal P}_t}{{\cal P}_{s}}= 16 c_s \epsilon .
\end{equation}
The observational data of Planck 2018 TT, TE, EE + LowE + Lensing + BK18 + BAO at the 95\% CL set an upper limit on the tensor-to-scalar ratio as $r \leq 0.036$ \cite{Paoleti:2022,bk18}.
The combination of Eqs. (\ref{eq:ntt}) and (\ref{eq:r}) gives the following consistency relation
\begin{equation}\label{eq:r-nt}
r =-8 c_s n_t.
\end{equation}
The number of inflationary $e$-folds from the end of inflation, in which $\phi=\phi_e$ and $\epsilon=1$, to the  CMB horizon crossing time, i.e., $\phi=\phi_i$  is given by
\begin{equation}\label{eq:N1}
N(\phi_i,\phi_e)=-\int_{\phi _e}^{\phi_i }\left ( \frac{H}{ \dot{\phi }} \right )\mathrm{d} \phi .
\end{equation}
Now we introduce the natural potential given by
\begin{equation}\label{eq:NPotential}
	V(\phi)=\Lambda^4\big[1+\cos(\phi/f)\big],
\end{equation}
for the non-canonical Lagrangian (\ref{Lagrangian}), wherein $\Lambda$ and $f$  are constant parameters with dimensions of mass. The observational predictions of the natural potential (\ref{eq:NPotential}) in the standard model of inflation are ruled out of the permitted domain of the Planck 2018 data \cite{akrami:2020}. Hence, we attempt to revive this potential in light of the latest observational data in the non-canonical framework, with the help of parameters $\alpha$ and $M$.

Note that in the non-canonical inflation, as stated in \cite{Baumann:2009}, only the equilateral shape of non-Gaussianity parameter is considered  as follows \cite{Rezazadeh:2015}
\begin{equation}\label{eq:fnl}
	f_{\rm NL}^{\rm equil}=-\frac{275}{972}\left ( \frac{1}{c_{s}^{2}}-1 \right )=-\frac{275}{486}\left ( \alpha -1 \right ) .
\end{equation}
The Planck measurements impose a constraint on the equilateral non-Gaussianity parameter as \cite{akrami:2018}
\begin{equation}\label{eq:fnlPlanck}
	f_{\rm NL}^{\rm equil}=-26\pm 47 .
\end{equation}
Equations (\ref{eq:fnl}) and (\ref{eq:fnlPlanck}) place an upper bound on the parameter $\alpha$  as $\alpha \leq 130$. Therefore, the variation of the non-canonical parameter $\alpha$ is considered within the range of integer values $2\leq\alpha\leq 130$.

At this stage, our aim is to solve the following equations to analyze the background evolution
\begin{equation}\label{eq:3equations}
P_s(\phi_i,M,\alpha,\Lambda,f)=2.1\times 10^{-9}, \quad \epsilon(\phi_e,M,\alpha,\Lambda,f)=1,  \quad N(\phi_i,\phi_e,M,\alpha,\Lambda,f)=N_0,
\end{equation}
which allows us to determine the value of the inflaton field at the CMB scale ($N=N_0$), signified by $\phi_i$, the value of the inflaton field at the end of inflation ($N=0$), denoted by $\phi_e$, and the non-canonical mass scale parameter $M$. To solve these three equations, the number of $e$-folds $N_0$ and the parameters $\alpha$, $f$ and $\Lambda$ must be specified.  Throughout this paper, we set $\Lambda=0.0075M_{\rm p}\simeq O(10^{16})~\rm Gev$  to achieve a permissable energy scale for inflation,  and the values of the other parameters are provided in the corresponding diagrams.

\begin{figure}[H]
\begin{minipage}[b]{1\textwidth}
\centering
\subfigure[\label{phi} ]{\includegraphics[width=0.45\textwidth]{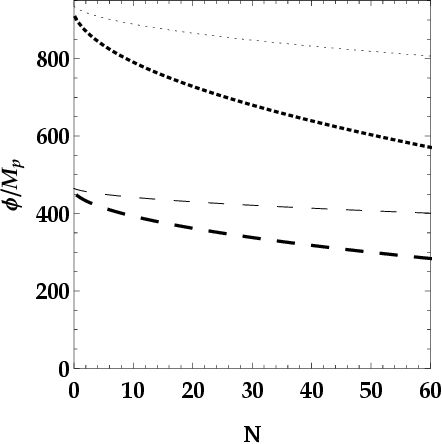}}\hspace{.1cm}
\subfigure[\label{h} ]{\includegraphics[width=0.45\textwidth]{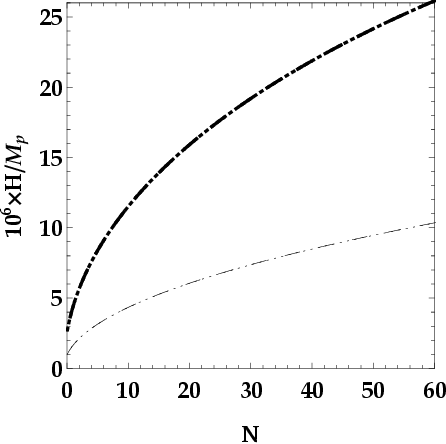}}
\subfigure[\label{eps} ]{\includegraphics[width=0.45\textwidth]{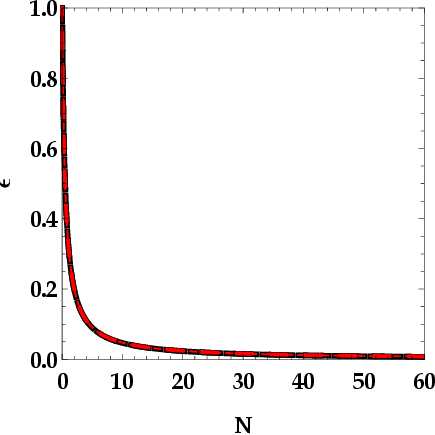}}\hspace{.1cm}
\subfigure[\label{eta} ]{\includegraphics[width=0.45\textwidth]{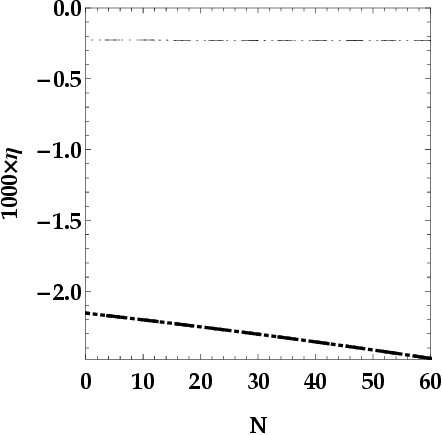}}
\end{minipage}
\caption{The evolutions of (a) inflaton field $\phi$, (b) Hubble parameter $H$, (c) the first slow-roll parameter $\epsilon$, and (d) the second slow-roll parameter $\eta$ as functions of the  $e$-folds number $N$. The dashed and dotted lines correspond to  $f=150M_p$ and $f=300M_p$, respectively, while the thick and thin lines represent $\alpha=2$ and $\alpha=80$, respectively. Note that in panel (c), to illustrate the overlap of the curves, $\alpha=2$ and $\alpha=80$ are shown by thick and thin red curves, respectively, confirming the perfect degeneracy.}
	\label{BG}
\end{figure}

It is notable that for some simple inflationary models, Eqs. (\ref{eq:3equations}) can be solved analytically. However, in the current study focused on non-canonical natural inflation, analytical solution is not feasible, and equations must instead be solved numerically. Once Eqs. (\ref{eq:3equations}) are solved, the resulting values of $\phi_i$, $\phi_e$ and $M$ are used to compute $\phi$, $H$, $\epsilon$ and $\eta$ as functions of the $e$-folds number $N$. The corresponding results are presented in Fig. \ref{BG}. Each panel in Fig. \ref{BG} includes four curves: thick and thin curves correspond to  $\alpha=2$ and $\alpha=80$, respectively, while dashed and dotted lines represent  $f=150M_p$ and $f=300M_p$, respectively. The diagram \ref{phi} shows that (i) the evolution of the inflaton field $\phi$ with respect to the $e$-folds number $N$ is influenced by both $\alpha$ and $f$ which indicates that there is no degeneracy in $\phi(N)$ with respect to these parameters; (ii) for a given $N$, increasing of the value of  $\alpha$ leads to a larger field value; (iii) the inflaton field increases during the inflationary era. It is inferred from Figs. \ref{h} and \ref{eta} that the evolution of the Hubble parameter $H$ and the second slow-roll parameter $\eta$ both depend on the value of $\alpha$. However, these quantities exhibit complete degeneracy with respect to the variation in $f$. This indicates that changes in $f$ do not significantly affect their evolution. Moreover, Fig. \ref{h} shows that the Hubble parameter $H$ decreases during the inflationary era and, for a given $N$, it increases as $\alpha$ decreases. Also, figure \ref{eps} confirms that (i) inflation ends at $N=0$ when $\epsilon=1$; (ii)  The first slow-roll parameter $\epsilon$ exhibits perfect degeneracy with respect to variations in both $\alpha$ and $f$. To illustrate this degeneracy, the cases $\alpha=2$ and $\alpha=80$ are represented by the thick and thin red curves, respectively, while dashed and dotted lines correspond to $f=150M_p$ and $f=300M_p$, respectively.

By substituting the solutions of Eqs. (\ref{eq:3equations}) into Eqs. (\ref{eq:epsilon V})-(\ref{eq:eta V}), and using the relevant expressions for the scalar spectral index $n_{\rm s}$ in Eq. (\ref{eq:inf_ns_SR}) and the tensor to scalar ration  $r$ in Eq. (\ref{eq:r}), one can obtain the values of these observables. In this analysis, we fix the parameters $N_0$ and $\alpha$ and solve Eqs. (\ref{eq:3equations}) for various values of $f$ within the interval $[2,316]M_p$ \cite{akrami:2020} to gain a parametric plot of $r$ versus $n_{\rm s}$.

For $\alpha=1$, the model reproduces the expected predictions of canonical natural inflation. Remarkably, for $\alpha\neq 1$, both $r$ and $n_{\rm s}$ exhibit complete degeneracy with respect to the parameter $f$. Also, for $\alpha \neq 1$,  $n_{\rm s}$ weakly depends on $\alpha$. For example with $\alpha=2$ and $N_0=50(60)$, varying $f$ across the full interval yields constant values of $r=0.077(0.066)$ and $n_{\rm s}=0.960(0.967)$. The same degeneracy behavior is observed for all other values of $\alpha$.

\begin{figure}[H]
	\centering
	\vspace{-0.2cm}
	\scalebox{1.3}[1.3]{\includegraphics{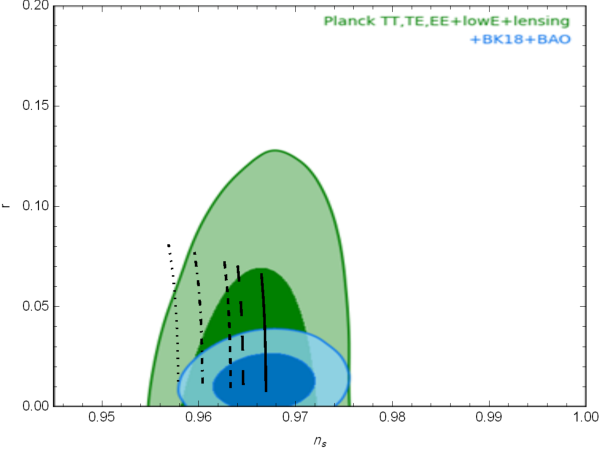}}
	\vspace{-0.1cm}
	\caption{The $r-n_{\rm s}$ diagrams for the non-canonical natural inflation for different $e$-folds number $N=60$ (solid curve), $N=56$ (long-dashed curve), $N=54$ (dashed curve), $N=50$ (dash-dotted curve) and $N=47$ (dotted curve). Along each curve, the parameter $\alpha$ varies in the range  $2\leq\alpha\leq 130$,  from  top to bottom. The dark (light) green area represents the 68\% (95\%) CL constraints from  Planck 2018 TT,TE, EE+low E+lensing data.  The dark (light) blue region shows the 68\% (95\%) CL constraints from the joint dataset Planck 2018 TT, TE, EE+low E+lensing+BK18+BAO. }
	\label{fig:0}
\end{figure}

The parametric behavior of $r$ versus $n_{\rm s}$ for several values of the $e$-folds number $N=(47,50,54,56,60)$ is plotted in Fig. \ref{fig:0} in light of the Plank$+$BICEP/Keck 2018 data. In this figure (i) along each curve, the parameter $\alpha$  varies from $2$ (top of the curve) to $130$ (bottom of the curve); (ii) the $r-n_{\rm s}$ curves indicate that $n_{\rm s}$ is nearly degenerate with respect to $\alpha$, while $r$ varies significantly with $\alpha$; (iii) the minimum duration of inflation based on the Planck 2018$+$BK18 data is $N=47\ (52)$ at the 95\%\ (68\%) CL.  Since the $r$ and $n_{\rm s}$ are perfectly  degenerate with respect to $f$, the value of $f$ is unconstrained by these plots. In other words, for any $f\in [2,316] M_P$, the same $r-n_{\rm s}$ curves are reproduced.

The permissable ranges of the parameter $\alpha$ for which the $(r-n_{\rm s})$ curves in Fig. \ref{fig:0} lie within the blue regions of the Planck 2018$+$BK18 data, are provided in Table \ref{tabalpha}.  This Table also presents the corresponding values of the non-canonical mass scale parameter $M$ in the allowed regions. Note that the  values of $M$ depend on $f$ which means that different values of $f$ yield different intervals for $M$. In this table we fix $f=2  M_P$, and the maximum value of $M$ corresponds to the smallest allowed value of $\alpha$. In other words, for fixed values of all other parameters,  $M$ is a decreasing function of $\alpha$.

\begin{table}[H]
  \centering
  \caption{Acceptable values of the non-canonical parameters $\alpha$ and the mass scale $\mu \equiv \frac{M}{M_p} $ corresponding to the allowed $e$-folds number $N$ in the context of the non-canonical natural inflation. These results are obtained by considering combined observational constraints from Planck and BICEP/Keck 2018 data on the  $(r - n_{\rm s})$ plane, equilateral non-Gaussianity ($f_{\rm NL}^{\rm equil}$), reheating parameters ($\omega_{\rm re}$, $N_{\rm re}$, $T_{\rm re}$), and relic gravitational waves. The potential parameter is fixed at $f = 2$.}
  \resizebox{\textwidth}{!}{\fontsize{6}{7}\selectfont%
  \begin{tabular}{|c|c!{\vrule width 0.2pt}c|c!{\vrule width 0.2pt}c|c!{\vrule width 0.2pt}c|c!{\vrule width 0.2pt}c|}
      \thickhline
      \multirow{3}{*}{$N$} &
      \multicolumn{4}{c|}{$(r - n_{\rm s}) + f_{\rm NL}^{\rm equil}$} &
      \multicolumn{4}{c|}{$(r - n_{\rm s}) + f_{\rm NL}^{\rm equil} + \omega_{\rm re} + N_{\rm re} + T_{\rm re} $} \\ \cline{6-9}
       & \multicolumn{4}{c|}{}
     & \multicolumn{4}{c|}{\makecell[c]{%
       $(r-n_{\rm s})+ f_{\rm NL}^{\rm equil}+ \omega_{\rm re}+N_{\rm re}+T_{\rm re}+ {\rm GWs}$}} \\ \cline{2-9}
      & \multicolumn{2}{c|}{95\% CL} & \multicolumn{2}{c|}{68\% CL} & \multicolumn{2}{c|}{95\% CL} & \multicolumn{2}{c|}{68\% CL} \\ \cline{2-9}
      & $\alpha$ & $\mu(\times 10^{-5})$ & $\alpha$ & $\mu(\times 10^{-5})$ & $\alpha$ & $\mu(\times 10^{-5})$ & $\alpha$ & $\mu(\times 10^{-5})$ \\
      \thickhline
      $47$ & $130$ & $26.84$ & $-$ & $-$ & $130$ & $26.84$ & $-$ & $-$ \\ \hline
      $50$ & $[16,130]$ & $[25.24,41.75]$ & $-$ & $-$ & $[16,130]$ & $[25.24,41.75]$ & $-$ & $-$ \\ \hline
      $54$ & $[9,130]$ & $[23.38,43.27]$ & $[23,130]$ & $[23.38,35.58]$ & $[9,130]$ & $[23.38,43.27]$ & $[23,130]$ & $[23.38,35.58]$ \\ \hline
      $56$ & $[7,130]$ & $[22.55,43.31]$ & $[17,130]$ &  $[22.55,36.70]$ & $[7,130]$ & $[22.55,43.31]$ & $[17,130]$ & $[22.55,36.70]$  \\ \hline
      $57$ & $[7,130]$ & $[22.16,42.49]$ & $[15,130]$ & $[22.16,37.02]$ & $-$ & $-$ & $-$ &$-$ \\ \hline
      $60$ & $[6,130]$ & $[21.05,40.93]$ & $[12,130]$ &  $[21.05,36.75]$ & $-$ & $-$ & $-$ & $-$ \\
      \thickhline
    \end{tabular}%
  }
  \label{tabalpha}
\end{table}

\section{Reheating epoch} \label{sec3}
At the end of inflation, the kinetic energy of the inflaton dominates, and the inflaton begins to oscillate around the minimum of the potential. This oscillatory phase is called the reheating epoch. Considering the smooth oscillatory behavior of the inflaton field around the minimum of the potential, the most consistent interpretation is that reheating proceeds through a perturbative decay of the inflaton into relativistic particles. Although the nature of reheating process is still not completely understood \cite{Baumann:2009}, considerations of this epoch  can be useful for testing inflationary models. In this epoch, the focus is on the reheating temperature ($T_{\rm re}$),  the reheating duration ($N_{\rm re}$) and the equation of state parameter ($\omega_{\rm re}$). Both the reheating temperature and  duration depend on the equation of state parameter. The number of $e$-folds during the reheating epoch must be positive, i.e., $N_{\rm re}\geq0$, and its value must be much smaller than the duration of the inflationary epoch. From
\cite{Martin2015} we have
\begin{equation}\label{eq:Nre}
	N_{\rm re}=\left(\frac{4}{1-3 \omega _{\rm re}}\right)\left [ -\frac{1}{4}\ln \left ( \frac{30}{\pi ^2g_{\rm re}} \right )-\frac{1}{3} \ln \left ( \frac{11 g_{\rm re}}{43} \right )-\ln \left ( \frac{k_\ast }{a_0 T_0} \right )-\ln \left ( \frac{\rho_e^{1/4}}{H_{\ast}} \right )-N\right ].
\end{equation}
In this equation, $k_\ast=0.05~{\rm Mpc^{-1}}$ is the pivot scale, $T_0=2.725~{\rm K}$ is the current temperature of the CMB radiation, $g_{\rm re}=106.75$ is the effective number of relativistic degrees of freedom, $a_0=1$ is the scale factor at the present time. Moreover, $\rho _e$ denotes the energy density of the inflaton field at the end of inflation, $N$ is the duration of inflation from the end of inflation to the horizon-crossing moment, and  $H_{\ast}$ is the Hubble parameter at the horizon-crossing time. By employing the  conversion factor $1~{\rm Mpc^{-1}}=6.39\times10^{-39}{\rm GeV}$, Eq. (\ref{eq:Nre}) can be rewritten as
\begin{equation}\label{eq:Nre1}
	N_{\rm re}=\frac{4}{1-3 \omega _{\rm re}}\left[61.55-\ln \left (\frac{\rho_e^{1/4}}{H_{\ast}} \right)-N\right],
\end{equation}
in which the value of $H_{\ast}$ can be determined from Eq. (\ref{eq:Ps-SR}) as follows
\begin{equation}\label{eq:Hi}
	H_{\ast}=\sqrt{ 8 \pi ^2 M_{\rm p}^2 c_s {\cal P}_s(k_\ast ) \epsilon }.
\end{equation}
From the definition of the equation of state parameter as
\begin{equation}\label{eq:Omega total}
	\omega_{\phi}\equiv \frac{p_\phi}{\rho_\phi}=-1+\frac{2}{3}\epsilon,
\end{equation}
and using  $\epsilon=1$ at the end of inflation, we have $\omega_{\phi_ e}=-\frac{1}{3}$, hence
\begin{equation}\label{eq:phi-rho1}
	p_e=-\frac{1}{3}\rho_e .
\end{equation}
Substituting Eqs. (\ref{eq:rho}) and (\ref{eq:Lp}) into Eq. (\ref{eq:phi-rho1}),  the energy density at the end of inflation  $\rho_e$ in the non-canonical framework with power-law Lagrangian (\ref{Lagrangian})  can be found as
\begin{equation}\label{eq:rho end}
	\rho _e=\left(\frac{3\alpha }{\alpha +1}\right)V_e ,
\end{equation}
in which $V_e=V(\phi_e)$ is the value of the inflationary potential at the end of inflation.

\begin{figure}[H]
	\centering
	\vspace{-0.2cm}
	\scalebox{1.3}[1.3]{\includegraphics{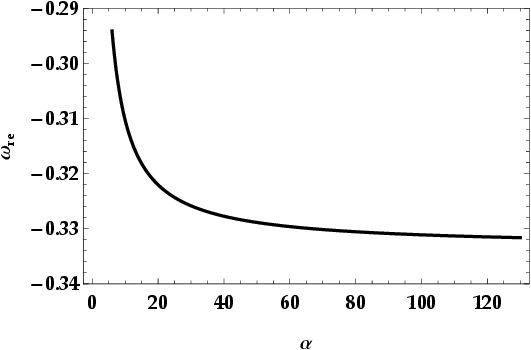}}
	\vspace{-0.1cm}
	\caption{Variation of the reheating equation of state parameter $\omega _{\rm re}$ with respect to the parameter $\alpha$ for $6\leq\alpha\leq 130$.}
	\label{fig:Omegare}
\end{figure}

\begin{figure}[H]
\begin{minipage}[b]{1\textwidth}
\centering
\subfigure[\label{Nre-ns}]{\includegraphics[width=0.45\textwidth]{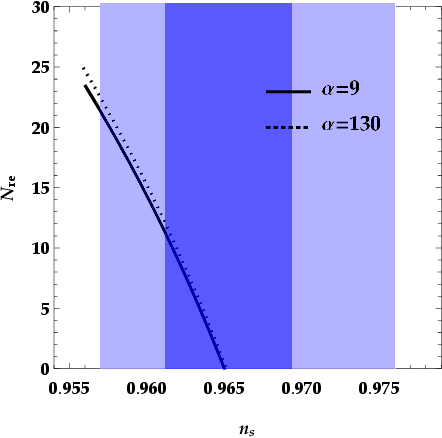}}\hspace{.1cm}
	\centering
	\vspace{-0.2cm}
\subfigure[\label{Nre-alpha} ]{\includegraphics[width=0.40\textwidth]{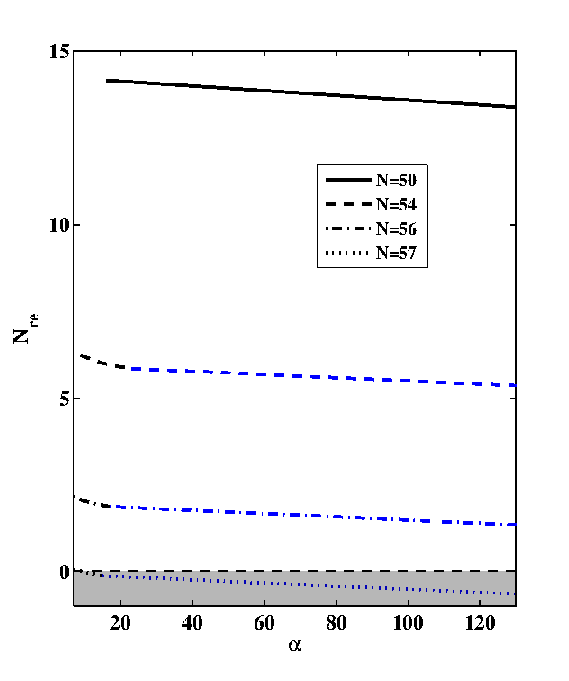}}
	\vspace{-0.1cm}
\end{minipage}
	\caption{The number of $e$-folds during reheating $N_{\rm re}$ versus (a) the scalar spectral index $n_{\rm s}$ for $\alpha=9$ (solid curve) and  $\alpha=130$ (dotted curve) and (b) the parameter $\alpha$ for $N=57$ (dotted curve), $N=56$ (dash-dotted curve), $N=54$ (dashed curve), and $N=50$ (solid curve). In panel (a), the dark and light blue regions  show the 68\% and 95\% CL, respectively, based on Planck 2018 TT,TE, EE + LowE + Lensing + BK18 + BAO data. In panel (b), the blue portions of each curve represent the 68\% CL allowed range for $\alpha$ corresponding to each $N$, as listed in Table \ref{tabalpha}, while the shaded areas mark the  regions excluded by $N_{\rm re}<0$.}
	\label{fig:1}
\end{figure}

According to \cite{Unnikrishnan:2012}, the reheating equation of state parameter is obtained as
\begin{equation}\label{action}
	1+\left \langle \omega _{\phi } \right \rangle=\left ( \frac{2\alpha }{2\alpha -1} \right )\left [ \int_{0}^{\phi _{m}}{\mathrm{d}\phi \left ( 1-\frac{V\left (\phi  \right ) }{V\left (\phi_{m}  \right )} \right ) }^{\frac{2\alpha -1}{2\alpha }}\right ]\left [\int_{0}^{\phi _{m}}{\mathrm{d}\phi \left ( 1-\frac{V\left (\phi  \right ) }{V\left (\phi_{m}  \right )} \right ) }^{\frac{-1}{2\alpha }}  \right ]^{-1} ,
\end{equation}
in which $\langle\omega_\phi\rangle$ denotes the mean value of the equation of state parameter over one oscillation period, and it is equal to the reheating equation of state parameter $\omega_{\rm re}$. In addition $\phi_m$ is the maximum value of the scalar field during this period. In the reheating epoch, assuming small amplitude oscillations of the scalar field around the minimum of potential ($\phi=\phi_{\rm{min}}$), the potential energy can be expanded in powers of ($\phi-\phi_{\rm{min}}$). In this expansion, the leading-order term  is proportional to $(\phi-\phi_{\rm{min}})^2$. By substituting this result into Eq. (\ref{action}), one can obtain
 \begin{equation}\label{omega}
	\omega _{\rm re}=	\left \langle \omega _{\phi } \right \rangle=\frac{1}{2\alpha-1}\left [1-\frac{\alpha\Gamma\left (\frac{3}{2}-\frac{1}{2\alpha}\right)}{\Gamma\left (\frac{5}{2}-\frac{1}{2\alpha}\right)} \right ],
\end{equation}
where $\Gamma$ denotes the gamma function. In Fig. \ref{fig:Omegare}, we plot variations of $\omega _{\rm re}$ versus the $\alpha$ parameter. Figure shows that as $\alpha$ increases from $1$ to $130$, the reheating equation of state parameter $\omega_{\rm{re}}$ decreases from $0$ to $-0.331$.
In addition, substituting Eqs. (\ref{eq:Hi}), (\ref {eq:rho end}), and (\ref {omega}) into Eq. (\ref {eq:Nre1}), the number of $e$-folds during reheating $N_{\rm re}$, is plotted as a function of the scalar spectral index in Fig. \ref{Nre-ns}, for $\alpha=9$ and $\alpha=130$. Since  $N_{\rm re}$ must be positive, we infer from Fig. \ref{Nre-ns} that, for $0.957\leq n_{\rm s}\leq0.965$, the duration of inflation should lie within the range $47\leq N\leq56$. Moreover, in Fig. \ref{Nre-alpha}, the reheating duration is plotted as a function of $\alpha$ for various values of the $e$-folds number. It can be inferred from this figure that (i) the degeneracy in model predictions with respect to $\alpha$ approximately persists even after including the effects of reheating, i.e., different values of $\alpha$ result in similar values for $N_{\rm re}$; (ii) the case $N=57$ is excluded, as it yields a negative value for  $N_{\rm re}$. Therefore, the minimum acceptable value for the number of inflationary $e$-folds is $N=56$.

\begin{figure}[H]
\begin{minipage}[b]{1\textwidth}
	\centering
	\vspace{-0.2cm}
\subfigure[\label{Tre-ns}]{\includegraphics[width=0.46\textwidth]{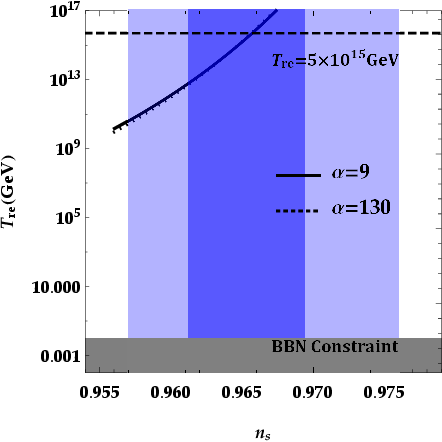}}\hspace{.1cm}
	\centering
	\vspace{-0.2cm}
\subfigure[\label{Tre-alpha} ]{\includegraphics[width=0.40\textwidth]{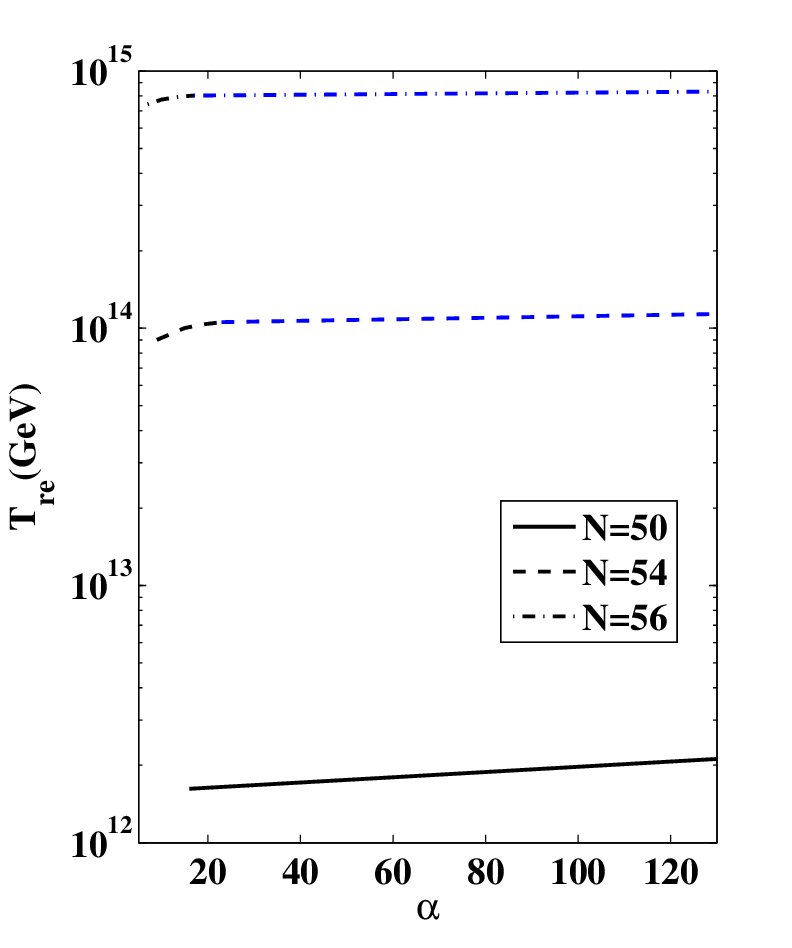}}
	
	\vspace{-0.1cm}
\end{minipage}
	\caption{Reheating temperature $T_{\rm re}$ versus (a) $n_{\rm s}$   and (b) $\alpha$ . In panel (a), the shaded gray area at the bottom indicates the region excluded by the BBN constraint, while the horizontal dashed line at the top marks the model-independent bound  $T_{\rm re} = 5\times 10^{15} GeV$. The dark (light) blue zones  are the same as in Fig. \ref{fig:1}. }
	\label{fig:2}
\end{figure}

Besides $N_{\rm re}$, the reheating temperature $T_{\rm re}$ is also used to further constrain inflationary models. It has been obtained in \cite{Martin2015} as
\begin{equation}
 T_{\rm re} =\Big( \frac{30 \rho_{\rm e}}{\pi^2 g_{\rm re}} \Big)^{\frac{1}{4}} e^{-\frac{3}{4}(1+\omega_{\rm re})N_{\rm re}}.\label{eq:Tre}
\end{equation}
The reheating temperature depends on both $\omega_{\rm re}$ and $N_{\rm re}$. Since, the reheating phase takes place between the end of inflation and the onset of  Big Bang Nucleosynthesis (BBN),  the reheating temperature is constrained by the following upper and lower bounds \cite{Kawasaki:2000,Hasegawa}
\beq
4 \, {\rm MeV} \leq T_{\rm re} \leq 5\times10^{15}\, {\rm GeV} .
\label{eq:bound_Tre}
\eeq
In Fig. \ref{Tre-ns}, the reheating temperature is plotted as a function of the scalar spectral index $n_{\rm s}$, for $\alpha=9$ (solid curve) and $\alpha=130$ (dotted curve). The vertical dashed line in this figure shows the boundary imposed by the condition $N_{\rm re}\geq0$. The model-independent upper bound, $T_{\rm re}=5\times 10^{15}{\rm GeV}$, is shown by a horizontal dashed line at the top of the panel, while the shaded gray area at the bottom refers to the region excluded by the BBN constraint. In Fig. \ref {Tre-alpha}, the reheating temperature is plotted versus $\alpha$ for $N=56$ (dash-dotted curve), $N=54$ (dashed curve), and $N=50$ (solid curve). It can be inferred from this figure that, even when incorporating the reheating temperature constraints,  the degeneracy of the model with respect to $\alpha$ remains, i.e., different values of $\alpha$ yield identical values for $T_{\rm reh}$.
It is worth noting that lattice simulations of reheating dynamics in \cite{Podolsky} suggest that the effective reheating equation of state parameter approach $\omega_{\rm eff}\sim 0.2-0.3$ once nonlinear preheating effects are included. In contrast, in the present non-canonical framework, $\omega_{\rm re}$ is derived analytically from Eq. (\ref{omega}) as a function of the parameter $\alpha$, for which $\omega_{\rm re}\in[-1/3,0]$ for $1\leq\alpha\leq 130$. This range corresponds to the mean equation of state parameter during the oscillatory phase when the potential behaves quadratically around its minimum. Adopting a larger effective $\omega_{\rm re}\sim 0.2-0.3$
in Eqs. (\ref{eq:Nre1}) and (\ref{eq:Tre}) would shorten the reheating duration and increase the temperature moderately, but our main conclusions regarding the allowed range of $N$, the weak dependence on $\alpha$, and perfect degeneracy with respect to $f$ would remain unchanged.
\section{Primordial gravitational waves} \label{sec4}
The propagation of primordial gravitational waves ensued from reentry of tensor modes into the horizon after inflation, is one of the most significant predictions of the inflationary scenario \cite{allen88}.
Due to their extremely weak interactions, GWs can preserve and transmit information from the early Universe across different cosmic epochs \cite{Bernal:2019lpc}.

The present-day energy density spectrum of these primordial GWs, particularly those re-entering the horizon during the reheating and RD eras, are related through the following expression \cite{Mishra:2021}
\begin{equation}
\Omega _{\rm GW_0}^{(\rm re)}(\nu)  = \Omega _{\rm GW_0}^{(\rm RD)}(\nu) \left (\frac{\nu}{\nu_{\rm re}}\right )^{2\left (\frac{\omega_{\rm re}-1/3}{\omega_{\rm re}+1/3}\right )},~~~ \nu_{\rm re} < \nu \leq \nu_{\rm e}.
\label{eq:GW_spectrum_2b}
\end{equation}
Here, $\Omega _{\rm GW_0}^{(\rm RD)}$ is the present-day GW density for modes re-entering the horizon during the RD era, and is given as a function of frequency $\nu$ by follows
\begin{equation}
\Omega _{\rm GW_0}^{(\rm RD)}(\nu)=\left ( \frac{1}{24} \right )r~{\cal P}_{\rm s} (k_{\ast })\left ( \frac{\nu}{\nu_{\ast }} \right )^{n_{{\rm t}}}\Omega _{\rm r_0},~~~\nu_{\rm eq}< \nu\leq \nu_{\rm re},
\label{eq:GWs_spectrum_2a}
\end{equation}
in which $\Omega_{\rm r_0}=2.47\times10^{-5}h^{-2}$ is the current radiation density parameter.
Here, $\nu_{\rm e}$, $\nu_{\rm re}$, $\nu_{\rm eq}$ and $\nu_{*}$ signify the frequencies of GWs corresponding to the end of inflation, end of reheating, matter-radiation equality, and CMB horizon crossing moment, respectively. Here, we set $\nu_{\rm e}=4.3\times 10^8 {~\rm Hz}$, $\nu_{\rm eq}=1.7\times 10^{-17}{~\rm Hz}$ and $\nu_{*}=2.4\times 10^{-16}{~\rm Hz}$.

The frequency of GWs can be formulated as a function of the comoving wavenumber $k$ as \cite{Mishra:2021}
\begin{equation}\label{fk}
 \nu=\frac{1}{2 \pi} \left( \frac{k}{a_0} \right),
\end{equation}
and can also be written as a function of the temperature as
\begin{equation}
\nu(T)=7.36\times 10^{-8} {~\rm Hz} \left ( \frac{g_{0}^{\rm s}}{g_{\rm T}^{\rm s}} \right )^{\frac{1}{3}}\left ( \frac{g_{\rm T}}{90} \right )^{\frac{1}{2}}\left ( \frac{T}{\rm GeV} \right),
\label{eq:GW_f}
\end{equation}
wherein  $g_{\rm T} = 106.75$ is the number of relativistic degrees of freedom in energy at temperature $T$, and $g_{0}^{\rm s} = 3.94$, $g_{\rm T}^{\rm s} = 106.75$ denote the effective relativistic degrees of freedom in entropy at the present and at temperature $T$, respectively.

\begin{figure}[H]
	\begin{minipage}[b]{1\textwidth}
		\centering
		\subfigure[\label{47} ]{\includegraphics[width=0.45\textwidth]{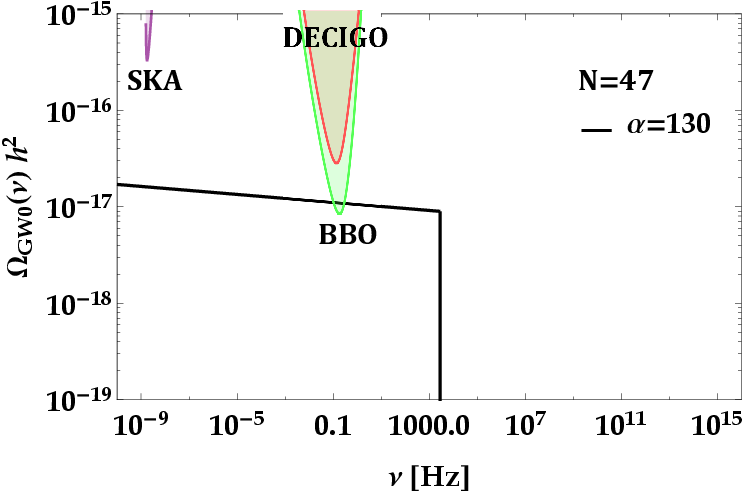}}\hspace{.1cm}
		\subfigure[\label{50} ]{\includegraphics[width=0.45\textwidth]{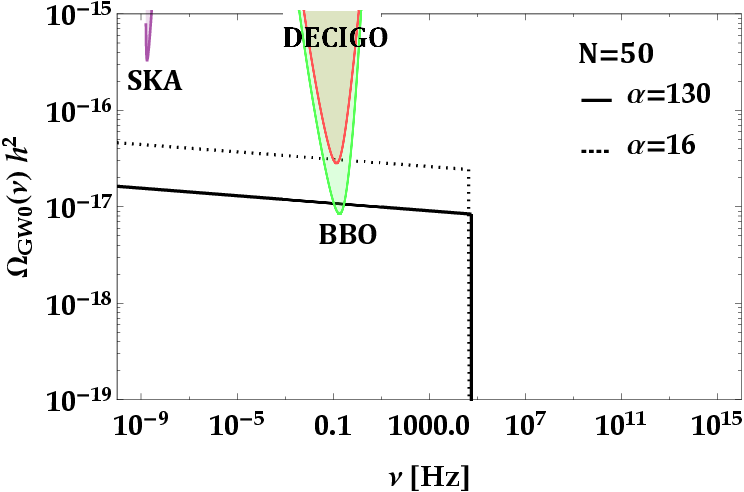}}
		\subfigure[\label{54} ]{\includegraphics[width=0.45\textwidth]{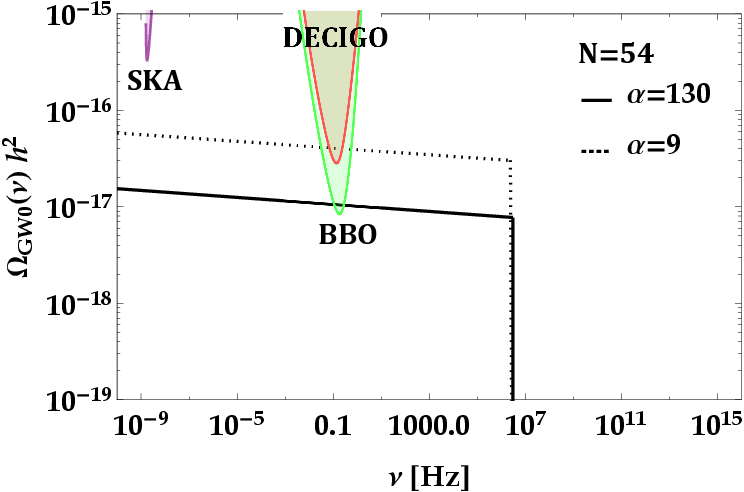}}\hspace{.1cm}
		\subfigure[\label{57} ]{\includegraphics[width=0.45\textwidth]{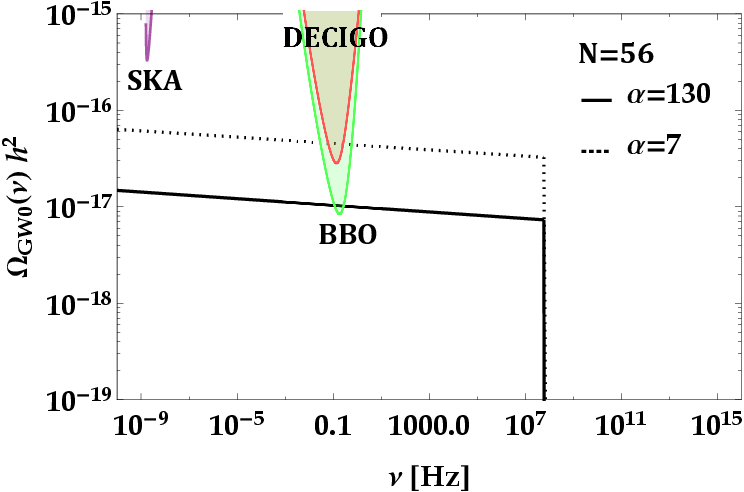}}
	\end{minipage}
	\caption{The present-day density GW spectrum as a function of the frequency for (a) $N=47$, (b) $N=50$, (c) $N=54$, and (d)  $N=56$. The solid (dotted) curves in each panel, correspond to the maximum (minimum) permitted values of $\alpha~(95\%~\rm CL$) according to Table \ref{tabalpha}.}
	\label{GW}
\end{figure}

Using  Eqs. (\ref{eq:GW_spectrum_2b})-(\ref{eq:GW_f}), the current density spectra of primordial GWs, $\Omega _{\rm GW_0}$, predicted by the model can be evaluated. Figure \ref{GW} illustrates $\Omega _{\rm GW_0}$ as a function of frequency for various values of $\alpha$ at specific inflationary  $e$-folds number, $N=(47,50,54,56)$. In each panel of this figure,
the GW spectra are compared with the sensitivity curves of future GW detectors such as BBO \cite{Corbin:2006BBO,Harry:2006BBO},
DECIGO \cite{Yagi:2011BBODECIGO,Yagi:2017BBODECIGO}, and SKA \cite{ska,skaWeltman:2020}. In each graph,
the solid (dotted) curve denotes the maximum (minimum) value of $\alpha$, for a particular $e$-folds number $N$ as listed in Table \ref{tab GW1}. It is inferred from Fig. \ref{GW} that (i) for each value of $N$, the predicted GW spectra lie within the sensitivity regions of  BBO and DECIGO, and they could be traceable in the future; (ii) for all cases, the GW spectrum corresponding to $\alpha=130$ falls within the sensitivity domain of BBO, while the spectrum related to the minimum value of $\alpha$ lies within the range detectable by DECIGO; (iii)  degeneracy of the model with respect to the parameter $\alpha$ is removed, using the inclusion of primordial GWs; (iv) the breaking point in each curve corresponds to the frequency at the end of the reheating, $\nu_{\rm re}=\nu(T=T_{\rm re})$, as obtained from Eq. (\ref{eq:GW_f}) (see Table \ref{tab GW1}).  It is important to note that, despite the degeneracy-breaking role of primordial GWs  with respect to $\alpha$, the model remains degenerate with respect to the parameter $f$. Specifically, for any value of $f\in [2,316] M_p$, the same results shown in Fig. \ref{GW} are reproduced. Moreover, the analysis of relic GWs does not further constrain the model parameters $f$ and $\alpha$, nor the number of inflationary $e$-folds.

In conclusion, by incorporating  joint constraints from Planck+BK18 data, reheating dynamics, and the prospective detection of primordial GW,  the viable range for  inflationary duration is determined to be $47\leq N\leq 56$ (95\% CL) and $52\leq N\leq 56$ (68\% CL) (see Table \ref{tabalpha}).

\begin{table}[H]
  \centering
  \caption{The reheating temperature $T_{\rm re}$ computed from Eq. (\ref{eq:Tre}), and corresponding frequency $\nu_{\rm re}=\nu(T=T_{\rm re})$ derived from Eq. (\ref{eq:GW_f}), for various values of the $e$-folds number $N$ and different values of the parameter $\alpha$. }
\resizebox{.6\textwidth}{!}{\scriptsize
\begin{tabular}{cccc}
\thickhline
$N$\qquad  &
\qquad$\alpha$\qquad &
\qquad $T_{\rm re}/{\rm GeV} $\qquad &
\qquad $\nu_{\rm re}/{\rm Hz} $\qquad  \\
\thickhline
\quad 47 \quad\qquad & 
\qquad$ 130 $ \qquad  & \qquad $1.07\times 10^{11}$\qquad & \qquad $29\times 10^2$  \\
\hline
\quad $\multirow{2}{*}{50} $\quad\qquad  & 
\qquad$16 $ \qquad &\qquad $1.62\times 10^{12}$ &\qquad $43\times 10^3$ \\
\quad &
\qquad$130$ \qquad &\qquad $2.12\times 10^{12}$ &\qquad $57\times 10^3$  \\
\hline
\quad $\multirow{2}{*}{54} $\quad\qquad  & 
\qquad$ 9 $ \qquad &\qquad $9.00\times 10^{13}$ &\qquad $2.40\times 10^6$ \\
\quad &
\qquad$130$ \qquad &\qquad $1.14\times 10^{14}$ &\qquad $3.03\times 10^6$  \\
\hline
\quad $\multirow{2}{*}{56} $\quad\qquad  & 
\qquad$7$ \qquad &\qquad $2.22\times 10^{15}$ &\qquad $5.93\times 10^7$  \\
\quad &
\qquad$130$ \qquad &\qquad $2.25\times 10^{15}$ &\qquad $6.02\times 10^7$  \\
\hline
\thickhline
\end{tabular}}
 \label{tab GW1}
\end{table}

\section{Conclusions}\label{sec5}
In this study, we investigated a non-canonical natural inflationary model with a power-law Lagrangian (\ref{Lagrangian}), which generalizes the standard canonical form. The observational predictions of the natural potential for the scalar spectral index $n_{\rm s}$ and the tensor-to-scalar ratio $r$ in the standard (canonical) framework are incompatible with the latest Planck 2018  and BICEP/Keck data. Moreover, within the canonical framework,  no degeneracies in  $n_{\rm s}$ and $r$ exist with respect to the model parameter $f$. We revived the natural potential in the non-canonical setting by employing the parameters $\alpha$ and $M$ in the power-law Lagrangian (\ref{Lagrangian}). However, in this framework degeneracies in  $n_{\rm s}$ and $r$ with respect to the parameters $f$ and $\alpha$ arise. Specifically,  $n_{\rm s}$  becomes independent of both  $f$ and $\alpha$, while $r$ becomes independent of $f$. In order to cure these degeneracies, we analyzed the reheating era and primordial GWs for our model. In particular, we computed the reheating parameters including the reheating equation of state parameter $\omega_{\rm re}$, duration $N_{\rm re}$, and temperature $T_{\rm re}$, to explore their dependency on the parameters $f$ and $\alpha$. Furthermore, we calculated the current energy spectrum of primordial GWs as a function of $\alpha$. The main conclusion of our study are summarized as follows

\begin{itemize}
  \item In the background evolution (Fig. \ref{BG}), the scalar field $\phi$ shows no degeneracy with respect to either $\alpha$ or $f$ (Fig. \ref{phi}), whereas the Hubble parameter $H$ and the second slow-roll parameter $\eta$ are degenerate only with respect to $f$ (Figs. \ref{h} and \ref{eta}). The evolution of $\epsilon$ exhibits perfect degeneracy with respect to both $\alpha$ and $f$ (Fig. \ref{eps}).
  \item The predictions of the non-canonical power-law natural inflationary model lie within the permitted region derived from Planck and BICEP/Keck 2018 data for specific intervals of the inflationary $e$-folds number $N$ and the non-canonical parameters $\alpha$ and $M$, as shown in Table \ref{tabalpha} and Fig. \ref {fig:0}.
  \item The predictions of the model for $n_{\rm s}$ and $r$ exhibit complete degeneracy with respect to $f$, while only $n_{s}$ is degenerate with respect to $\alpha$.
  \item The minimum duration of inflation was determined as $N=47$ (52) at 95\% (68\%) CL, based on constraints of Planck and BICEP/Keck data on the $r-n_{\rm s}$ plane.
  \item The reheating equation of state parameter $\omega_{\rm reh}$ was obtained analytically as a function of  $\alpha$ through Eq. (\ref{omega}). For $6\leq\alpha\leq 130$, we found  $0\leq\omega_{\rm reh}\leq -0.331$ (see Fig. \ref{fig:Omegare}).
  \item Applying the constraint $N_{\rm re}\geq 0$, we obtained an upper bound on  the inflationary duration of  $N=56$. Therefore,  the viable range for inflationary $e$-folds number is $47\leq N\leq 56$.
   \item Using the reheating constraints on $N_{\rm re}$ and $T_{\rm re}$ do not fully remove the  degeneracies  with respect to  $\alpha$ and $f$ (see Figs. \ref{Nre-alpha}-\ref{Tre-alpha}).
  \item The predicted density spectrum of primordial GWs for $N=(47,50,54,56)$ and various values of  $\alpha$ lies within the sensitivity zones of future  GW detectors such as BBO and DECIGO. This enables the model to be probed observationally.
  \item The GW spectrum is sensitive to $\alpha$. This provides a means to resolve the degeneracy with respect to this parameter. However, the degeneracy with respect to $f$ remains, and resolving it is left for future works.
\end{itemize}
On the whole, natural inflation in the non-canonical power-law framework is reconciled with present observational constraints for specific ranges of $N$ and $\alpha$. The analysis of reheating era and primordial GWs offers additional ways to constrain  inflationary models and address degeneracies. This emphasizes the importance of integrated observational and theoretical limitations in refining the inflationary models.

\subsection*{Acknowledgements}
The authors thank the referees for their valuable comments and Milad Solbi for valuable discussions.


\end{document}